\documentclass[reprint,amsmath,amssymb,aps,twocolumn,superscriptaddress]{revtex4-1}

\UseRawInputEncoding

\usepackage[nopar]{lipsum}
\usepackage{graphicx}
\usepackage{dcolumn}
\usepackage{bm}
\usepackage[capitalize]{cleveref}

\usepackage{siunitx}

\begin{document}
\preprint{APS/123-QED}

\title{Tunable quantum interferometer for correlated moir\'e electrons}




\author{Shuichi Iwakiri}
\thanks{siwakiri@phys.ethz.ch; amestre@phys.ethz.ch; \\ These two authors contributed equally.}
\author{Alexandra Mestre-Tor\`a}
\thanks{siwakiri@phys.ethz.ch; amestre@phys.ethz.ch; \\ These two authors contributed equally.}
\author{El\'ias Portol\'es}
\affiliation{Laboratory for Solid State Physics, ETH Zurich,~CH-8093~Zurich, Switzerland}
\author{Marieke Visscher}
\affiliation{Faculty of Science, Leiden Institute of Physics (LION), Leiden University, Rapenburg 70, 2311 EZ Leiden, Netherlands}
\author{Marta Perego}
\author{Giulia Zheng}
\affiliation{Laboratory for Solid State Physics, ETH Zurich,~CH-8093~Zurich, Switzerland}
\author{Takashi Taniguchi}
\affiliation{Research Center for Materials Nanoarchitectonics, National Institute for Materials Science,  1-1 Namiki, Tsukuba 305-0044, Japan}
\author{Kenji Watanabe}
\affiliation{Research Center for Electronic and Optical Materials, National Institute for Materials Science, 1-1 Namiki, Tsukuba 305-0044, Japan}
\author{Manfred Sigrist}
\affiliation{Institute for Theoretical Physics, ETH Zurich,~CH-8093~Zurich, Switzerland}
\author{Thomas Ihn}
\author{Klaus Ensslin}
\affiliation{Laboratory for Solid State Physics, ETH Zurich,~CH-8093~Zurich, Switzerland}
\affiliation{Quantum Center, ETH Zurich,~CH-8093 Zurich, Switzerland}


\begin{abstract}
Magic-angle twisted bilayer graphene (MATBG) can host an intriguing variety of gate-tunable correlated states, including superconducting and correlated insulator states. Junction-based superconducting devices, such as Josephson junctions and SQUIDs, have been introduced recently and enable the exploration of the charge, spin, and orbital nature of superconductivity and the coherence of moir\'e electrons in MATBG.
However, complementary fundamental coherence effects --- in particular, the Little--Parks effect in a superconducting and the Aharonov--Bohm effect in a normal conducting ring --- remained to be observed.
Here, we report the observation of both these phenomena in a single gate-defined ring device where we can embed a superconducting or normal conducting ring in a correlated or band insulator. We directly observe the Little--Parks effect in the superconducting phase diagram as a function of density and magnetic field, confirming the effective charge of $2e$.
By measuring the Aharonov--Bohm effect, we find that in our device, the coherence length of normal conducting moir\'e electrons exceeds a few microns at 50 mK. Surprisingly, we also identify a regime characterized by $h/e$-periodic oscillations but with superconductor-like nonlinear transport. Taken together, these experiments establish a novel device platform in MATBG, and more generally in tunable 2D materials, to unravel the nature of superconductivity and other correlated quantum states in these materials.
\end{abstract}

\maketitle

\newpage

\twocolumngrid

Magic-angle twisted bilayer graphene (MATBG) with its moir\'e flat band \cite{suarez_morell_flat_2010,zhang_nearly_2019} constitutes a condensed-matter system to realize a wide variety of correlated states, such as superconducting and correlated insulator states, that are tunable by gating \cite{cao_correlated_2018,cao_unconventional_2018,yankowitz_tuning_2019,lu_superconductors_2019,lu_superconductors_2019,saito_independent_2020,stepanov_untying_2020,di_battista_revealing_2022}.
A novel class of gate-defined nanodevices, including Josephson junctions \cite{de_vries_gate-defined_2021,rodan-legrain_highly_2021} and SQUIDs \cite{portoles_tunable_2022},  have been recently realized in MATBG. These structures have provided excellent platforms for controlling mesoscopic superconductivity and characterizing MATBG.
Extending this approach to a doubly-connected geometry without any junction, namely a ring, promises unique microscopic information about the material and the device.

The physical properties of a ring threaded by a magnetic field are in general periodic in flux quanta \cite{byers_theoretical_1961,bloch_josephson_1970} $\Phi_0=\frac{h}{e^*}$, with $e^{*}$ being the charge of the carrier.
In a superconducting ring ($e^{*}=2e$), $h/2e$-periodic oscillations of critical temperature and critical current appear. These oscillations are known as the Little--Parks effect and were the first experimental evidence for the $2e$ charge pairing in conventional superconductors \cite{little_observation_1962,groff_fluxoid_1968}. 
In fact, the Little--Parks effect can be used to determine the charge of the superconducting carriers\cite{balents_superconductivity_2020,yang_plethora_2023}, complementing the Josephson-junction and SQUID experiments\cite{de_vries_gate-defined_2021, portoles_tunable_2022}. 
Moreover, properties of unconventional superconductors can be revealed by anomalies of the Little--Parks effect, such as a phase shift \cite{liu_chiral_2018,hua_theory_2022,aoyama_little-parks_2022,geshkenbein_vortices_1987,xu_spin-triplet_2020,li_observation_2019,almoalem_evidence_2022} or a change in periodicity \cite{loder_magnetic_2008,juricic_restoration_2008,zhu_magnetic_2010, fernandes_charge-4e_2021,jiang_charge-4e_2017, almoalem_evidence_2022}, and thereby help to understand the underlying superconducting symmetry.

By contrast, a normal conducting ring  ($e^{*}=e$) shows $h/e$-periodic oscillations of resistance, the Aharonov--Bohm effect, and works as a direct probe to quantify the phase coherence of electrons. Given the low Fermi velocity and the large effective mass in MATBG, a possible non-Fermi liquid nature of its flat band electrons \cite{cao_unconventional_2018,cao_strange_2020}, and intrinsic disorders introduced by twist-angle inhomogeneity \cite{uri_mapping_2020}, quantifying the phase coherence length is key to understanding the dynamics of moir\'e electrons.
In addition, the phase coherence length enables the estimation of the penetration depth of a superconducting wave function into the normal conducting state (proximity effect) \cite{de_gennes_boundary_1964,clarke:jpa-00213516}, which plays an important role in gate-defined superconducting devices.
However, the exploration of these fundamental quantum interference effects has been hampered by the lack of a suitable device architecture and the sensitivity of the moir\'e superlattice to disorder \cite{uri_mapping_2020}, which poses a challenge to the conventional approach of fabricating a ring by physical/chemical etching.

\begin{figure*}[t!]
\includegraphics[width=1\textwidth]{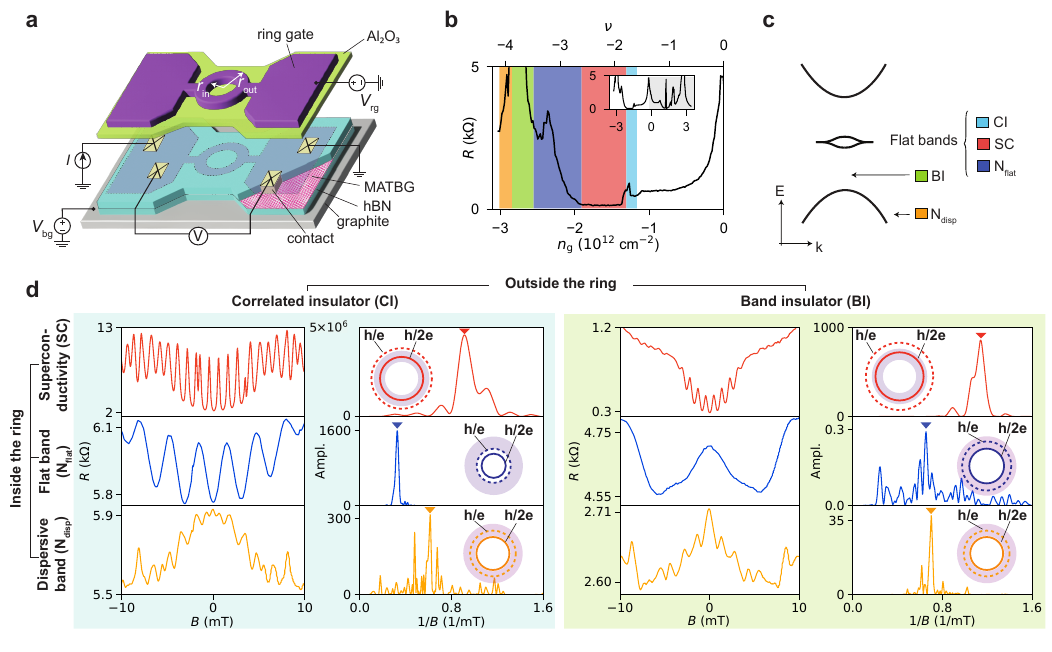}
\caption{\textbf{Highly tunable quantum interference.
a} Device and measurement schematics. The graphite--hBN--MATBG--hBN heterostructure is contacted by four electrodes (yellow). The ring gate (purple) is formed on top of an aluminium oxide layer (green). The lithographic inner ($r_\textrm{in}=\SI{600}{nm}$) and outer ($r_\textrm{out}=\SI{1000}{nm}$) radius of the ring gate are indicated. DC-voltage sources are connected to the ring gate and back gate. A four-terminal measurement is performed by applying a constant current and measuring the voltage drop across the ring. \textbf{b} Resistance of the MATBG as a function of carrier density $n_\textrm{g}$ at $V_\textrm{rg} = \SI{0}{V}$. The regions with coloured background correspond to the density ranges of the states introduced in panel \textbf{c}. The inset shows the resistance for both the electron (grey background) and hole side (white background).
\textbf{c} Band structure schematics of MATBG. The colour-coded labels indicate the correlated insulator (CI), superconducting (SC), normal conducting in the flat band ($\text{N}_\text{flat}$), normal conducting in the dispersive band ($\text{N}_\text{disp}$) and band insulator (BI) states, respectively. \textbf{d} Overview of quantum-interference effects. A ring-shaped conducting path is defined by surrounding the ring with a CI (left panel, blue-shaded) or BI (right panel, green-shaded) state. For both insulating states outside the ring, the magneto-resistance oscillations and their FFT spectrum are shown for the three conducting states: SC (top row), $\text{N}_\text{flat}$ (middle row) and $\text{N}_\text{disp}$ (bottom row). The triangle in the spectrum marks the peak frequency. The insets show the effective radius of the ring for both $h/2e$ and $h/e$ oscillations (solid and dashed circles, respectively), on top of the lithographic area of the ring gate (purple).}
\label{SampleAndExampleTraces}
\end{figure*}

Here we present such an architecture with a gate-defined ring consisting of a loop that can be tuned to be superconducting or normal conducting, surrounded by a correlated or band insulator.
We confirm $2e$ pairing via the Little--Parks effect, and show that the phase coherence length of moir\'e electrons surpasses several microns at 50 mK, evidenced by $h/e$-periodic Aharonov--Bohm oscillations. We also discover an intriguing regime in which $h/e$-periodic oscillations appear alongside superconductor-like transport.
These results highlight the promise of the novel quantum interferometer in MATBG for studying interference phenomena of exotic quantum states of 2D materials.


\textbf{Highly tunable quantum interference}

We develop the gate-defined ring architecture shown in Fig.~\ref{SampleAndExampleTraces}a. We base the design on the proof-of-principle device reported in ref. \cite{iwakiri_2022} using Bernal bilayer graphene. The MATBG is encapsulated in hexagonal boron nitride (hBN) and is contacted by four electrodes. The sample is dual-gated with a graphite back gate and a metallic ring-shaped top gate (ring gate).
We operate the ring by first tuning the back gate voltage $V_\textrm{bg}$, which affects the entire MATBG area and induces a global density $n_\textrm{g}$. Then we tune the density under the ring gate, $n_\textrm{r}$, via the voltage $V_\textrm{rg}$. 
The ring has a lithographic inner radius of $r_\textrm{in}=\SI{600}{nm}$ and an outer radius of $r_\textrm{out}=\SI{1000}{nm}$. 
Through the electrodes, the sample is biased with a current $I$, and the voltage drop $V$ is measured in a four-terminal configuration.
Unless stated otherwise, the measurements are performed in a $^3$He--$^4$He dilution refrigerator at a temperature of $\SI{50}{mK}$.

The resistance $R$ of the MATBG (Fig.~\ref{SampleAndExampleTraces}b) as a function of $n_\textrm{g}$ with $V_\textrm{rg}$ set to zero shows pronounced peaks at charge neutrality, and in the correlated insulator (CI) and band insulator (BI) regimes on the hole side ($n_\textrm{g} < 0$). From the density at the BI peak, we estimate an average twist angle of $1.1 ^{\circ}$. 
When tuning the density beyond the correlated insulator, we observe superconductivity (SC) through a resistance drop.
We also access two normal conducting regimes: one inside the flat band (N$_\textrm{flat}$) and the other in the dispersive band (N$_\textrm{disp}$). Figure~\ref{SampleAndExampleTraces}c summarizes the relevant quantum states that form in the device.
Further details of the experimental setup are given in the Methods and Extended Data sections.

In order to define a conducting path, we tune the ring-shaped region into the SC, N$_\textrm{flat}$, or N$_\textrm{disp}$ regimes.
This conducting path is then surrounded by either CI or BI states to confine electrons.
As we will see below, the insulating state does not influence the observed interference pattern but has an effect on the quantum state distribution across the structure (see \ref{fig:simulation}).
Figure~\ref{SampleAndExampleTraces}d shows the resistance ($R$) oscillations in perpendicular magnetic field ($B$) for the six regimes at zero bias current and a temperature of 150 mK, together with their fast Fourier transform (FFT) spectra. 
When calculating the FFT spectrum, we subtract a smooth background extracted with the Savitzky--Golay filter. We convert the peak of the spectrum into an area assuming either $h/e$ or $h/2e$-periodicity as the relevant flux quantum and then compare the result with the lithographic radius of the ring (see inset circles in Fig.~\ref{SampleAndExampleTraces}d).

\begin{figure*}[t!]
\centering
\includegraphics{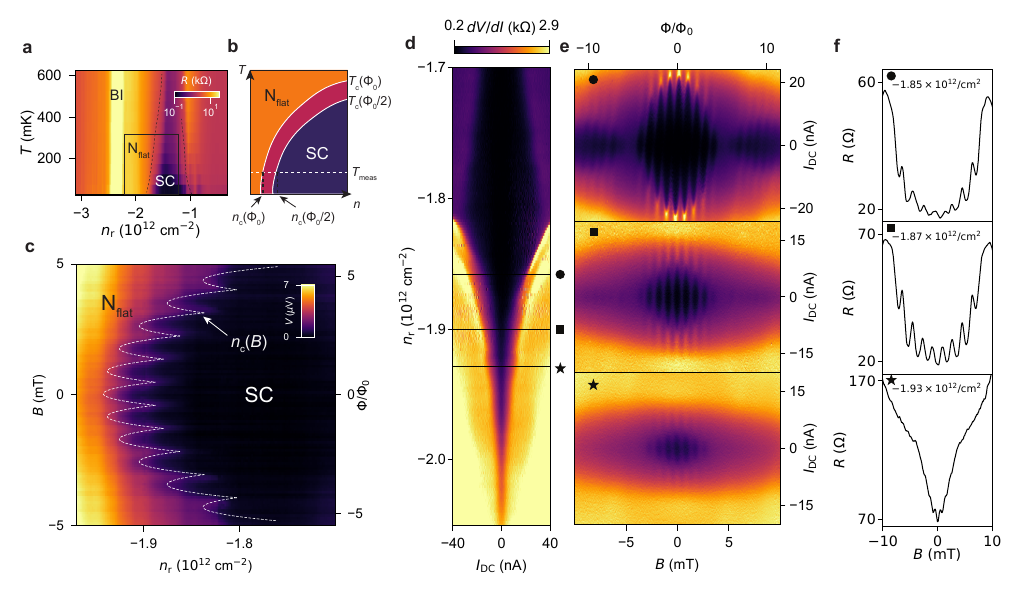}
\caption{\textbf{Tunable Little--Parks oscillations. a} Phase diagram of the device in the (SC, BI) configuration. The SC, N$_\textrm{flat}$ and BI states are represented. The square indicates the ($T$, $n$) domain for which we sketch the schematics in panel \textbf{b}. \textbf{b} Schematics of the superconducting dome in the phase diagram of MATBG. $T_\textrm{c} (\Phi_0)$ and $T_\textrm{c} (\Phi_0/2)$ mark the boundary between the N$_\textrm{flat}$ (orange) and SC (pink when $\Phi = n \Phi_0$,  and purple when $\Phi = \frac{n}{2} \Phi_0$) phases. Upon applying a magnetic field, the phase diagram breathes between $T_\textrm{c} (\Phi_0)$ and $T_\textrm{c} (\Phi_0/2)$. \textbf{c} Voltage drop $V$ across the ring as a function of $n_\textrm{r}$ and $B$, at $I = \SI{5.0}{nA}$. The white dashed curve (a guide to the eye) marks the phase boundary $n_\textrm{c}(B)$ between the SC and N$_\textrm{flat}$ states. \textbf{d} $\frac{dV}{dI}$ as function of $n_\textrm{r}$ and $I_\textrm{DC}$ when defining the ring in the BI state with $n_\textrm{g} = -2.94 \times 10^{12}\SI{}{\cm }^{-2}$. \textbf{e} Magneto-resistance oscillation of the critical current taken at the densities indicated in \textbf{d} with a black circle, square, and star, respectively. \textbf{f} Magneto-resistance oscillations taken at $I_\textrm{DC}=0$ nA and at the same densities as in \textbf{e}.
}
\label{TunableLP}
\end{figure*}

In the case of a superconducting ring (top row in Fig.~\ref{SampleAndExampleTraces}d, $n_\textrm{r}=-1.89 \times 10^{12}\SI{}{\cm }^{-2}$), the frequency peak appears at 0.92/mT for (SC, CI) and at 1.20/mT for (SC, BI). Hereafter, we denote the state inside and outside the ring as (inside, outside).
Assuming $h/2e$-periodicity, the observed frequency peaks correspond to an effective radius 
$r_\textrm{eff}$ of $\SI{767}{nm}$ and $\SI{855}{nm}$, respectively. These values are comparable to the center-line radius of the ring gate $r_\textrm{mid}= \frac{r_\textrm{in} + r_\textrm{out}}{2} = \SI{800}{nm}$. In contrast, the $r_\textrm{eff}$ when assuming $h/e$-periodicity does not match the lithographic dimension of the ring gate ($r_\textrm{eff}>\SI{1000}{nm}=r_\textrm{out}$). 
In these regimes, we further observe critical current and critical density oscillations, as we discuss in Fig.~\ref{TunableLP}.
Based on our findings, we attribute these oscillations to the $h/2e$-periodic Little--Parks effect, confirming that the charge of the superconducting carrier is $2e$.

When the ring is tuned into the dispersive band (bottom row in Fig.~\ref{SampleAndExampleTraces}d, $n_\textrm{r}=-3.58 \times 10^{12}\SI{}{\cm }^{-2}$), $h/e$-periodic oscillations appear with a spectrum covering a significant range in $1/B$.
The peak frequency of the oscillations is 0.621/mT for both (N$_\textrm{disp}$, CI) and (N$_\textrm{disp}$, BI). This frequency is approximately half of those observed in the superconducting case, suggesting an $h/e$-periodicity. The effective radii both assuming $h/e$- ($r_\textrm{eff}=\SI{873}{nm}$) and $h/2e$-periodicity ($r_\textrm{eff}=\SI{617}{nm}$) fit within the ring dimensions.
However, the oscillation amplitude decays exponentially in temperature and strives at higher magnetic fields than the Little--Parks oscillations (see \ref{Ext_RawRB_andTdep}). The amplitude decay in temperature is characteristic of the Aharonov--Bohm oscillations due to the smearing of the Fermi function and the reduction of the phase coherence length \cite{hansen_mesoscopic_2001}. We therefore attribute the oscillations to the $h/e$-periodic Aharonov--Bohm effect.
The measurement of the 
temperature dependence also allows us to estimate the phase coherence length $L_{\varphi}$ of electrons in the dispersive band (see \ref{Ext_RawRB_andTdep}).
At 50 mK, the coherence length is $\sim12.3\pm0.3$ $\SI{}{\micro \meter}$ in (N$_\textrm{disp}$, CI) and $\sim18.7\pm1.0$ $\SI{}{\micro \meter}$ in (N$_\textrm{disp}$, BI). These values exceed the perimeter of the ring ($2\pi r_\textrm{eff}\simeq \SI{4.80}{\micro m}$). This result demonstrates that the phase coherence of moir\'e electrons in the dispersive band is well preserved despite several sources of disorder such as twist-angle inhomogeneity and strain distribution.

\begin{figure}[t!]
\centering
\includegraphics{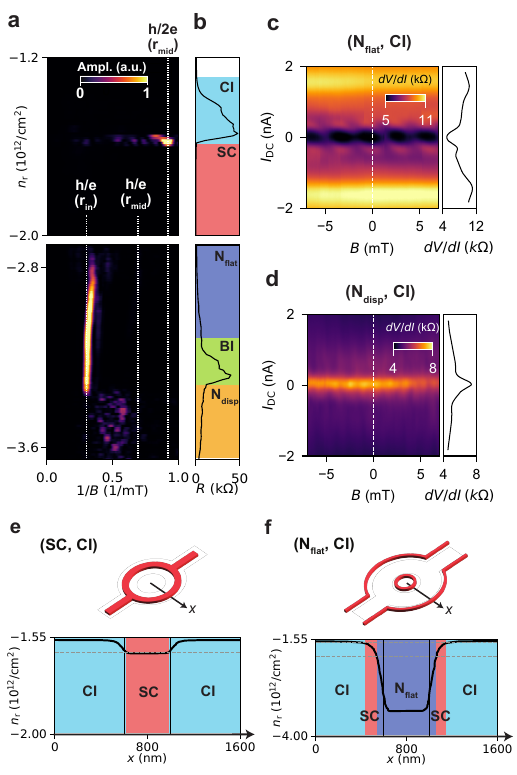}
\caption{\textbf{Spectroscopy of the quantum oscillations.}  \textbf{a} FFT spectrum of the magneto-resistance oscillations when sweeping the state inside the ring, with CI as the insulating state. The white lines in the spectrum indicate the range of frequencies expected for $h/e$ and $h/2e$ oscillations, taking the inner radius ($r_\textrm{in}$) and mean radius ($r_\textrm{mid}$) of the ring. \textbf{b} Resistance as a function of $n_\textrm{r}$, the different quantum states (CI, SC, N$_\textrm{flat}$, BI and N$_\textrm{disp}$) are indicated.
\textbf{c,d} $\frac{dV}{dI}$ as a function of $B$ and $I_\textrm{DC}$ in (N$_\textrm{flat}$, CI) regime (\textbf{c}) and (N$_\textrm{disp}$, CI) regime (\textbf{d}). In the side panel, linecut of $\frac{dV}{dI}$ at $B=0$. \textbf{e,f} Electrostatic simulation of the carrier density and quantum state distribution in the (SC, CI) regime (\textbf{e}) and (N$_\textrm{flat}$, CI) regime (\textbf{f}). The illustrations at the top show the superconducting regions (red) on the lithographic structure of the ring (black). At the bottom, plot of the spatial distribution of the carrier density along the radial axis of the ring (arrow in the illustration). The black vertical lines indicate the lithographic dimensions of the ring (400 nm width). Different quantum states (CI, SC, and N$_\textrm{flat}$) are attributed depending on the density $n_\textrm{r}$.}
\label{Nflat}
\end{figure}

Furthermore, there is a striking contrast between the oscillation in the normal conducting flat band regimes depending on the surrounding insulators (middle row in Fig.~\ref{SampleAndExampleTraces}d, $n_\textrm{r}=-3.10 \times 10^{12} \SI{}{\cm }^{-2}$).
In the (N$_\textrm{flat}$, CI) regime, we observe an oscillation with a frequency of 0.630/mT, even lower than in the SC and N$_\textrm{disp}$ cases. The effective radius assuming $h/2e$ periodicity is $r_\textrm{eff}=\SI{432}{nm}$, even smaller than $r_\textrm{in}$, while $r_\textrm{eff}=\SI{605}{nm}$ for $h/e$-periodicity matches $r_\textrm{in}$. Though the geometric argument points towards $h/e$-periodicity, this regime exhibits superconductor-like transport as well. We discuss this point in more detail later in Fig. \ref{Nflat}.
On the other hand, in the (N$_\textrm{flat}$, BI) regime, we observe magneto-resistance oscillations with very small amplitude. In this regime, the frequency peak appears close to the one in the $N_\textrm{disp}$ regime. The effective radius is $r_\textrm{eff}=\SI{813}{nm}$ assuming $h/e$-periodicity and $r_\textrm{eff}=\SI{585}{nm}$ assuming $h/2e$-periodicity. This regime shows neither nonlinear transport nor a drop in resistance with temperature. Therefore, we attribute the oscillations in the (N$_\textrm{flat}$, BI) regime to $h/e$-periodic Aharonov--Bohm oscillations. The phase coherence length, estimated from the temperature dependence, is $L_{\varphi}\sim6.51\pm1.32$ $\SI{}{\micro \meter}$, which is by a factor of 2--3 smaller than that for N$_\textrm{disp}$. This relatively short $L_{\varphi}$ in the flat band can be attributed to a large electron effective mass. In fact, the phase coherence length is proportional to the Fermi velocity $v_\textrm{Fermi}$, which depends inversely on the effective mass $m^{*}$ ($L_{\varphi}\propto v_\textrm{Fermi}\propto 1/m^{*}$), and the measured $m^{*}$ in the flat and dispersive bands are different by a factor of 1--10.\cite{cao_unconventional_2018} 

These results demonstrate that one can switch between the Little--Parks and the Aharonov--Bohm effects of MATBG by gate tuning. They also reveal the $2e$ charge pairing and long coherence lengths of the moir\'e electrons. In the following sections, we discuss each regime in more detail.

\textbf{Tunable Little--Parks oscillations}

The Little--Parks effect is essentially the magneto-oscillation of the free energy of the superconducting state\cite{tinkham2004introduction}, which results in the oscillation of the critical temperature $T_\textrm{c}$ and the critical current $I_\textrm{c}$. While the critical current is readily measurable, the current injection inevitably drives the system out of equilibrium, possibly giving rise to unwanted effects such as local breakdown of superconductivity \cite{fink_quantum-interference_1987,fink_superconducting_1988,moshchalkov_quantum_1993}. Therefore, a measurement at equilibrium is preferable. Besides, measurement of the $T_\textrm{c}$ oscillation is experimentally challenging as the expected amplitude is in the sub-mK range \cite{tinkham2004introduction}.
Here, taking advantage of the in-situ tunability of the carrier density in MATBG, we demonstrate an alternative route to detect Little--Parks oscillations. We probe the oscillation of $T_\textrm{c}$ by translating it into the oscillation of the critical density $n_\textrm{c}$ at which the superconducting transition occurs, enabling the detection of the Little--Parks effect near equlibrium.
Figure~\ref{TunableLP}a shows the phase diagram of the device in the (SC, BI) regime, measuring the resistance as a function of temperature $T$ and $n_\textrm{r}$ at zero magnetic field. Increasing the temperature from 50 mK to 600 mK, the density range of the SC state shrinks, forming a superconducting dome.
%
Due to the Little--Parks effect, the $T_\textrm{c}$ of the superconducting ring oscillates with the magnetic flux. As depicted in Fig.~\ref{TunableLP}b, this results in a compression and expansion of the superconducting phase boundary.
For a fixed temperature $T_\textrm{meas}$, such breathing can be translated into an oscillation of the critical density $n_\textrm{c}$, between $n_\textrm{c}(\Phi = \Phi_0)$ and $n_\textrm{c} (\Phi = \frac{1}{2} \Phi_0)$. It is therefore possible to probe the magneto-oscillation of the phase boundary by fixing the temperature and sweeping the carrier density.

Figure~\ref{TunableLP}c shows the measured voltage drop $V$ across the ring as a function of $n_\textrm{r}$ and $B$, at a DC bias current of $\SI{5}{nA}$ and in a density range close to the high-density edge of the superconducting dome (see \ref{fig:LPTdep} for the temperature and current dependence of the map).
At a fixed $B$, a jump from zero to a finite voltage $V$ marks the transition from the superconducting to the normal conducting regime. The magnetic field dependence of the density $n_\textrm{c}(B)$ represents the phase boundary. At $B=0$, $n_\textrm{c}$ is the maximum in absolute value. The shape of this boundary oscillates with a period of 0.870 mT, agreeing with an $h/2e$-periodicity ($r_\textrm{eff} \approx r_\textrm{mid}$). The smooth shift of the phase boundary with increasing $B$ to higher electron densities reflects the shrinking of the superconducting dome in $B$.
We observe the same result with a reversed current ($\SI{-5}{nA}$).
All in all, this means that MATBG exhibits a conventional Little--Parks effect, unlike the shifted pattern observed in some unconventional superconductors \cite{liu_chiral_2018,almoalem_evidence_2022}.
The demonstration of the Little--Parks effect near equilibrium by constructing the phase diagram is one of the distinct advantages of the gate-defined architecture.

We can also track the development of the oscillations as the state departs from the vicinity of the SC-to-N$_\textrm{flat}$ transition into deep inside the superconducting dome by tuning the density $n_\textrm{r}$.
Fig.~\ref{TunableLP}d shows $dV/dI$ as a function of DC current $I_\textrm{DC}$ and $n_\textrm{r}$ at zero magnetic field, and Fig.~\ref{TunableLP}e the $B$- and $I_\textrm{DC}$-dependence of $dV/dI$ for $n_\textrm{r}=-1.85,-1.87 \textrm{ and } -1.93 \times 10^{12} \SI{}{cm}^{-2}$ (see solid horizontal lines in Fig.~\ref{TunableLP}d).
The periodic oscillations of the critical current $I_\textrm{c}$ emerge on top of a decreasing background in magnetic field. The periodicity of the $I_\textrm{c}$ oscillations is $h/2e$ (0.87 mT), in agreement with the (SC, BI) data in Fig. \ref{SampleAndExampleTraces}d.
As $n_\textrm{r}$ is tuned to a value that shows larger $I_\textrm{c}$ (deep inside the superconducting dome), the amplitude of the $I_\textrm{c}$ oscillations increases and they are also observed up to higher magnetic fields. Moreover, the magnetic field origin of the oscillations shifts from zero and also depends on the direction of the applied current (see Fig. \ref{PhaseShift}), which can be due to the inductance of the ring.
At the lowest density (Fig. \ref{TunableLP}e, top panel), we observe that the critical current vanishes at $\sim 5$ mT and re-emerges at a higher magnetic field. This pattern resembles the Fraunhofer pattern of Josephson junctions.
However, as we discuss in the Supplementary Information section, the width of this hypothetical Josephson junction does not fit any of our ring geometries. We therefore attribute the observed pattern rather to the existence of multiple interference paths within the ring, generating a beating pattern.

The differential resistance traces at zero bias current ($I_\textrm{DC}=0$ nA) for the three densities are shown in Fig. \ref{TunableLP}f. Close to the SC-to-N$_\textrm{flat}$ transition (bottom panel, star symbol), the magneto-resistance oscillations are barely visible as the difference of resistance between SC and the normal states is vanishing. When the density is increased (middle panel, square symbol), we observe pronounced oscillations where a chain of parabolas appears on top of a smooth parabolic background. When the density is further increased (top panel, circle symbol), the oscillation amplitude drops again. This is because the state is deep in the superconducting dome and zero current is too remote from the transition to fully capture the Little--Parks effect.
As a matter of fact, deeper in the superconducting dome, one can only observe the critical current oscillations.


\textbf{Spectroscopy of the quantum oscillations}

Figure \ref{Nflat}a shows the magneto-oscillation spectrogram as a function of the density $n_\textrm{r}$ when the state outside the ring is CI (see Fig.~\ref{SampleAndExampleTraces}d, left column). The spectrogram is constructed by first measuring the differential resistance at zero bias current sweeping the magnetic field between $\pm$ 20 mT. Then the FFT spectrum is calculated after subtracting the smooth background from the raw data. In the spectrogram, we observe several regions with prominent peaks in the FFT. The resistance $R$ at zero magnetic field and corresponding quantum states are shown in Fig.~\ref{Nflat}b. 

An $h/2e$ peak is observed ($r_\textrm{eff}=r_\textrm{mid}$) at the edge of the SC region in the vicinity of the CI regime. This can be understood by the fact that the Little--Parks oscillation only appears at the onset of the superconducting transition.
When the density is tuned further down to the ($N_\textrm{flat}$, CI) regime, a prominent peak at 0.313 /mT appears between $-3.40 \times 10^{12}\SI{}{\cm }^{-2}<n_\textrm{r}<-2.80 \times 10^{12}\SI{}{\cm }^{-2}$. This corresponds to an effective radius of $r_\textrm{eff}=\SI{605}{nm}$ assuming $h/e$-periodicity, and $r_\textrm{eff}=\SI{432}{nm}$ assuming $h/2e$. As the latter $r_\textrm{eff}$ is considerably smaller than $r_\textrm{in}$, we attribute an $h/e$-periodicity to the oscillations. The density window across which this peak extends is unexpectedly wide and includes the BI regime, meaning that the oscillations persist even when the ring is mostly insulating. When entering the (N$_\textrm{disp}$, CI) regime, the spectrum becomes broader and features a peak within the $h/e$-periodic range ($r_\textrm{in}<r_\textrm{eff}<r_\textrm{mid}$). This broad spectrum is not unexpected for conventional Aharonov--Bohm oscillations as the aspect ratio of the gate-defined ring is small ($r_\textrm{in}/r_\textrm{out}\simeq0.6$), making possible many interference paths with different effective enclosed areas.
The oscillations in (N$_\textrm{flat}$, CI) and in (N$_\textrm{disp}$, CI) not only differ in peak frequency and frequency extent but also show different responses to $I_\textrm{DC}$. Figure~\ref{Nflat}c,d show the $I_\textrm{DC}$ and $B$ mapping in (N$_\textrm{flat}$, CI), with $n_\textrm{r}=3.07\times10^{12}$cm$^{-2}$, and (N$_\textrm{disp}$, CI), with $n_\textrm{r}=3.51\times10^{12}$cm$^{-2}$ respectively.
Interestingly, a superconductor-like behaviour of the differential resistance (dip at $I_\textrm{DC}=0$ nA) is observed in the (N$_\textrm{flat}$, CI) regime, despite the high resistance reaching up to a few k$\Omega$. In addition, a periodic chain of low-resistance states appears by sweeping $I_\textrm{DC}$ and $B$.
By contrast, the (N$_\textrm{disp}$, CI) regime exhibits a cusp in differential resistance at $I_\textrm{DC}=0$ nA, and no characteristic pattern is observed in the $I_\textrm{DC}$ and $B$ mapping. This again supports the interpretation that Aharonov--Bohm oscillations are observed for (N$_\textrm{disp}$, CI).


We attribute the superconducting behaviour of the (N$_\textrm{flat}$, CI) regime to the presence of a small residual superconducting region, emerging from the smooth evolution of the carrier density from inside to outside of the ring. To support this idea, we perform an electrostatic simulation of the carrier density distribution (see Supplementary Information). As shown in Fig. \ref{Nflat}e, in (SC, CI) regime, SC is the only state inside the ring. However, when the ring is detuned from the (SC, CI) to the (N$_\textrm{disp}$, CI) regime (Fig. \ref{Nflat}f), the spatial distribution of the quantum states becomes complex and a superconducting region can be embedded around the ring. This spurious superconducting region consists of a small ring with radius $\sim \SI{400}{nm}$ and a path surrounding the gated region, as the schematics of Fig. \ref{Nflat}f depicts.

Further improvement of the simulation, taking the proximity effect at the interface of different quantum states into account, can help to understand this regime better.


\textbf{Conclusion}

In conclusion, we present a gate-defined quantum interferometer in MATBG that provides a versatile platform for investigating the quantum coherence and the charge of correlated electrons from superconducting to normal conducting regimes.
We observe the Little--Parks effect by constructing the superconducting phase diagram as well as by measuring the oscillation of the magneto-resistance and critical current, confirming the charge-$2e$ pairing. We also observe the Aharonov--Bohm effect for the dispersive and flat band electrons in the same device. From it, we extract the phase coherence length exceeding a few microns, highlighting the robustness of its phase coherence.
We find a regime that exhibits magneto-resistance and critical current oscillations even detuned from the superconducting regime, which might be due to the electrostatic constriction of the device.

These experiments demonstrate that exploring the Little--Parks effect has the potential to provide insight into both the charge and spin nature of the Cooper pair in MATBG. Notably, the measurement under an in-plane magnetic field along with the perpendicular field could enable the observation of phase shifts in Little--Parks oscillations. Such shifts could serve as indicators of unconventional pairings such as spin-triplet superconductivity\cite{aoyama_little-parks_2022,yasui_little-parks_2017,jang_observation_2011}.
Moreover, the gate-defined architecture and the measurement scheme presented here can generally be implemented in other 2D superconductors (e.g., twisted multilayer graphene\cite{cao_pauli-limit_2021}, Bernal bilayer graphene\cite{zhou_isospin_2022}, and bilayer graphene/transition metal dichalcogenide\cite{zhang_enhanced_2023}), opening up the path towards the direct quantification of charge, spin, and coherence of correlated electrons in a plethora of exotic quantum states.

\begin{acknowledgements}
We thank Peter M\"{a}rki, Thomas B\"{a}hler, and the staff of the ETH cleanroom facility FIRST for technical support. 
We thank Lev Ginzburg and Rebekka Garrais for their help with the experiments.
We thank Andreas Trabesinger for his valuable input during the writing process of this manuscript.
We acknowledge financial support by the European Graphene Flagship Core3 Project, H2020 European Research Council (ERC) Synergy Grant under Grant Agreement 951541, the European Union’s Horizon 2020 research and innovation program under grant agreement number 862660/QUANTUM E LEAPS, the European Innovation Council under grant agreement number 101046231/FantastiCOF, NCCR QSIT (Swiss National Science Foundation, grant number 51NF40-185902).
K.W. and T.T. acknowledge support from the JSPS KAKENHI (Grant Numbers 21H05233 and 23H02052) and World Premier International Research Center Initiative (WPI), MEXT, Japan.
E.P. acknowledges support of a fellowship from ”la Caixa” Foundation (ID 100010434) under fellowship code LCF/BQ/EU19/11710062.
\end{acknowledgements}


\setcounter{figure}{0}
\renewcommand{\thefigure}{Supplementary Information\arabic{figure}}
\renewcommand{\theequation}{S\arabic{equation}}
\setcounter{secnumdepth}{2}

\setcounter{figure}{0}
\renewcommand{\thefigure}{Supplementary Information \arabic{figure}}
\renewcommand{\theequation}{S\arabic{equation}}
\setcounter{secnumdepth}{2}

\newpage
\clearpage
\onecolumngrid
\clearpage
\onecolumngrid

\clearpage

\section{Supplementary Information}
\subsection{Measurement setup}
We carry out all the measurements in a dilution refrigerator that uses a mixture of $^3$He and $^4$He with a base temperature of $\SI{55}{mK}$.
We apply a constant current bias between a pair of contacts across the ring and measure the voltage drop between another pair, also across the ring. To generate the bias current, we use an in-house-built d.c. source in series with a $\SI{100}{M\Omega}$ resistor. We use a d.c. amplifier, also built in-house, and measure its output with a Hewlett Packard 3441A digital multimeter. Each gate is connected to a different voltage source of the same type as the one used for generating a d.c. current. We convert the voltages we apply to the gates to electron densities by a parallel plate capacitor model. We estimate the capacitance per unit area of the back and ring (top) gate to be $C_\textrm{bg} = \varepsilon_0 \varepsilon_\textrm{hBN} / d_\textrm{bot}$ and $C_\textrm{rg} = \varepsilon_0 \varepsilon_\textrm{hBN} \varepsilon_\textrm{AlOx} / (\varepsilon_\textrm{hBN} d_\textrm{top} + \varepsilon_\textrm{AlOx} d_\textrm{AlOx})$, where $\varepsilon_0$ is the vacuum permittivity, $\varepsilon_\textrm{hBN}$ and $\varepsilon_\textrm{AlOx}$ are the relative permittivities of the hBN and the aluminium oxide, $d_\textrm{top}$ and $d_\textrm{bot}$ are the thicknesses of the top and bottom hBN and $d_\textrm{AlOx}$ is the thickness of the aluminium oxide layer. We calculate the electron density of the bulk as $n_\textrm{g} = C_\textrm{bg} V_\textrm{bg} / e$ and of the region below the top gate as  $n_\textrm{r} = \left(C_\textrm{bg} V_\textrm{bg} + C_\textrm{rg} V_\textrm{rg}\right) / e$, where $e$ is the elementary charge.

\subsection{Twist angle estimation}
We extract the twist angle of the sample using the relation $\theta = 2 \arcsin \left( \frac{a}{2L} \right)$  \cite{cao_correlated_2018}. In this expression, $a$ is the lattice constant of graphene and $L$ is the moir\'e periodicity, which represents the distance between two adjacent AA-stacked regions. In turn, $L$ is related to the area $\mathcal{A}$ of the moir\'e unit cell via $L = 2 \sqrt{2 \mathcal{A}/\sqrt{3}}$. Within a moir\'e unit cell, four electrons can be accommodated due to spin and valley degeneracy. Then, the band insulator peak due to the twist appears at the electron density $n_\textrm{BI}$ which corresponds to the occupation of 4 electrons per moir\'e unit cell $\mathcal{A} = \frac{4}{n_{\textrm{BI}}}$. We obtain $n_{\textrm{BI}}$ from the Landau fan and density mapping (\ref{fig:S_D}). Our analysis yields to an approximate twist angle of $1.11^\circ$.

\subsection{Device fabrication}
The device stack is assembled using the dry pick-up method~\cite{kim_vdw_2016}. We exfoliate graphene and hexagonal boron nitride (hBN) flakes on a $\SI{285}{nm}$ $\mathrm{p:Si/SiO_2}$ wafer. We start by scratching a graphene flake in two using a tungsten needle with a tip diameter of $\SI{2}{\micro m}$ controlled by a micromanipulator. We pick up all the flakes using a polydimethylsiloxane/polycarbonate stamp. We first pick up the top hBN flake, with a thickness of $\SI{18}{nm}$, at $\SI{90}{^\circ C}$. Then we proceed to pick up the graphene and assemble the twisted structure. For this, we first pick up half of the pre-cut graphene, rotate  the microscope stage by $1.1^\circ$ and then pick up the other half of the graphene, all at $\SI{40}{^\circ C}$. We encapsulate the graphene by picking up a bottom hBN flake, with a thickness of $\SI{55}{nm}$. For the encapsulation, the stack is first contacted to the bottom hBN at $\SI{40}{^\circ C}$ and the temperature of the stage is raised to $\SI{80}{^\circ C}$. Finally, we pick up a graphite flake of $\SI{29}{nm}$ at $\SI{100}{^\circ C}$ that serves as a back gate. The stack is then deposited at $\SI{160}{^\circ C}$ on a $\mathrm{p:Si/SiO_2}$ chip. After deposition, we clean the polycarbonate stamp using dichloromethane.

We contact the MATBG with edge contacts made by electron beam lithography followed by reactive ion etching, using $\mathrm{CHF_3/O_2}$ (40/4 sccm, 60W). The contacts are then evaporated using Cr/Au (10/80 nm). We define the electrode lines in two steps using electron beam lithography and depositing Cr/Au (10/60 nm for the first step and 5/50 nm for the second). Then we etch the stack to define the mesa and deposit a \SI{20}{nm} thick layer of aluminium oxide by atomic layer deposition. For the ring-shaped top gate, we again use electron beam lithography and evaporate Cr/Au (5/35 nm). The electrode line for the top gate is also done by electron beam lithography and evaporation using Cr/Au (10/70 nm).

\newpage
\subsection{Optical images and design of the sample}
\begin{figure}[h]
\centering
\includegraphics{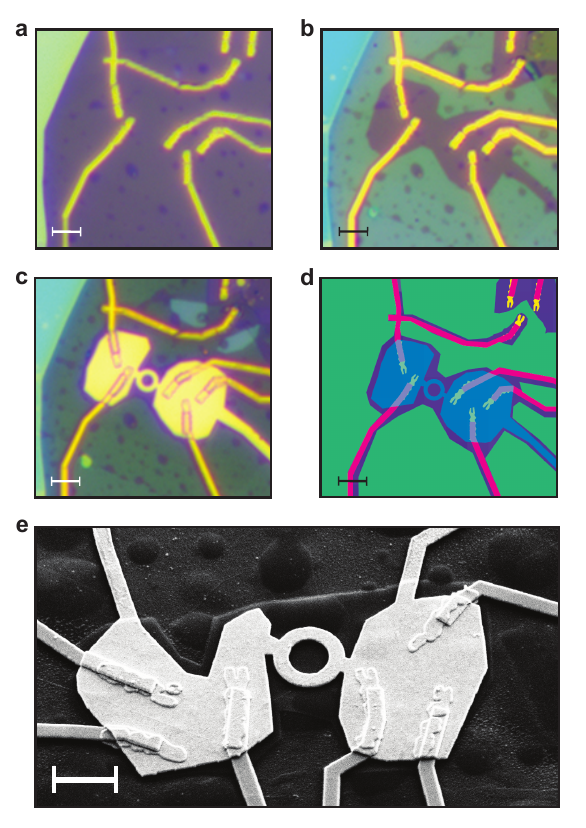}
\caption{The device used in this work is designed and fabricated through a series of steps summarized in the Methods section. \textbf{a} Microscope picture of the device after forming contacts, \textbf{b} after etching the mesa and \textbf{c} after depositing a layer of aluminum oxide and the ring gate. \textbf{d} Design of the device. Contacts are represented in yellow, etched areas in green, and the ring gate in blue (half transparent). The scale bar in \textbf{a}-\textbf{d} is \SI{5}{\micro m}. \textbf{e} SEM picture of the dummy device (not the one used in the measurement). The scale bar is \SI{2}{\micro m}.}
\label{Supp_sample}
\end{figure}

\clearpage
\newpage
\subsection{Raw data for magneto-resistance}
\begin{figure}[h]
\centering
\includegraphics[scale=0.85]{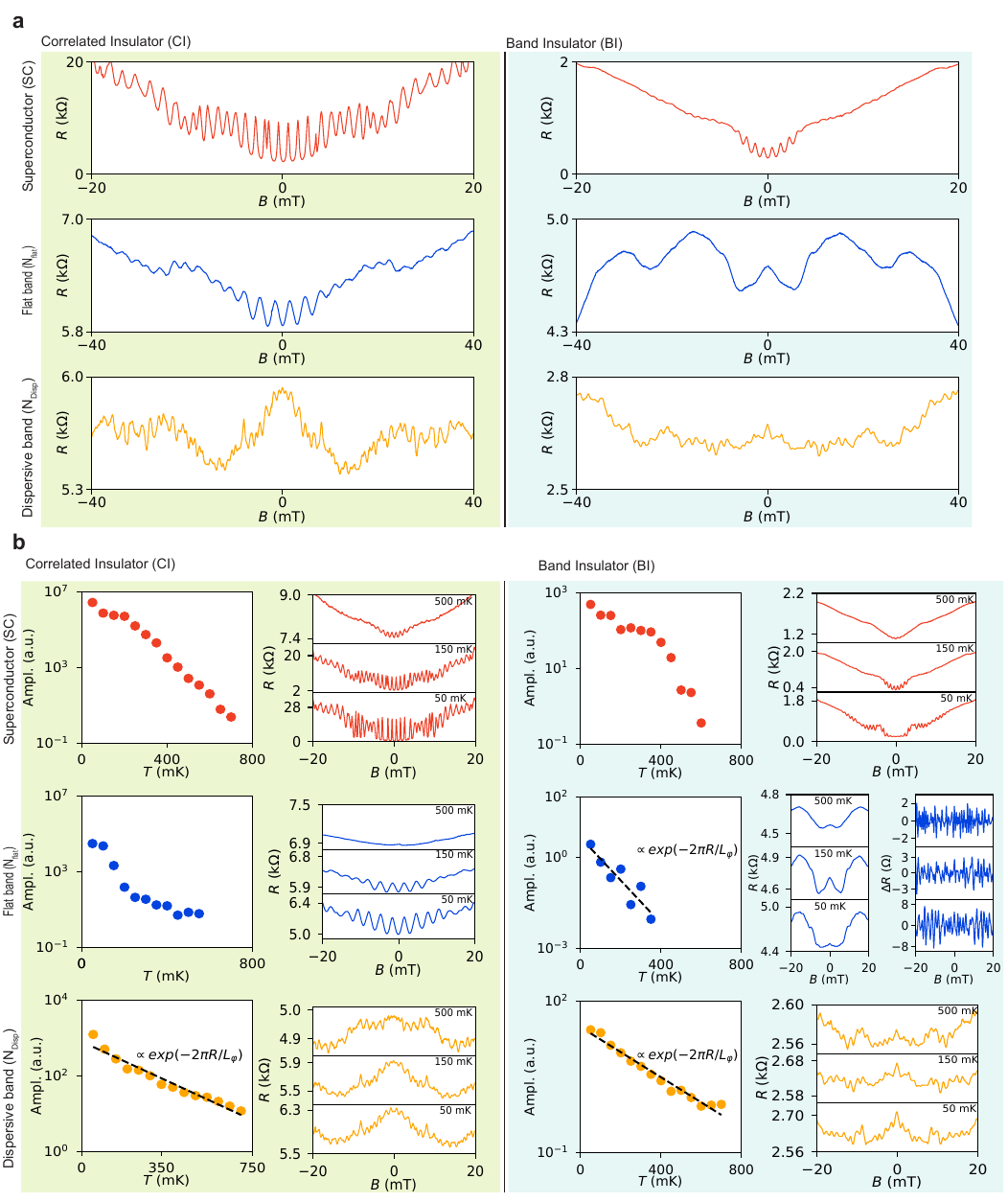}
\caption{\textbf{a} Magneto-resistance traces discussed in Fig. \ref{SampleAndExampleTraces} without background subtraction (raw data). \textbf{b} Temperature dependence of the oscillations. The gate voltages for each configuration are the same as the ones in Fig. \ref{SampleAndExampleTraces}. For the Aharonov--Bohm oscillations in (N$_\textrm{disp}$, CI) and (N$_\textrm{disp}$, BI) regimes, the temperature dependence is fitted with an exponential function $\textrm{e}^{-2\pi r_\textrm{eff}\alpha T}$, with $r_\textrm{eff}$ being the effective radius of the ring estimated by the FFT peak frequency and $\alpha$ is the coefficient that defines the temperature dependence of the phase coherence length $L_{\phi}=1/\alpha T$. From this fitting, we obtain $L_{\phi}\sim12.3$ $\pm$ 0.8 $\SI{}{\micro \meter}$ for (N$_\textrm{disp}$, CI) regime, $L_{\phi}\sim6.5$ $\pm$ 1.3 $\SI{}{\micro \meter}$ for (N$_\textrm{flat}$, BI), and $L_{\phi}\sim$18.7$\pm$1.0 $\SI{}{\micro \meter}$ for (N$_\textrm{disp}$, BI) regime.}
\label{Ext_RawRB_andTdep}
\end{figure}

\clearpage
\newpage
\subsection{Density maps and angle extraction}
\begin{figure}[h!]
\centering
\includegraphics{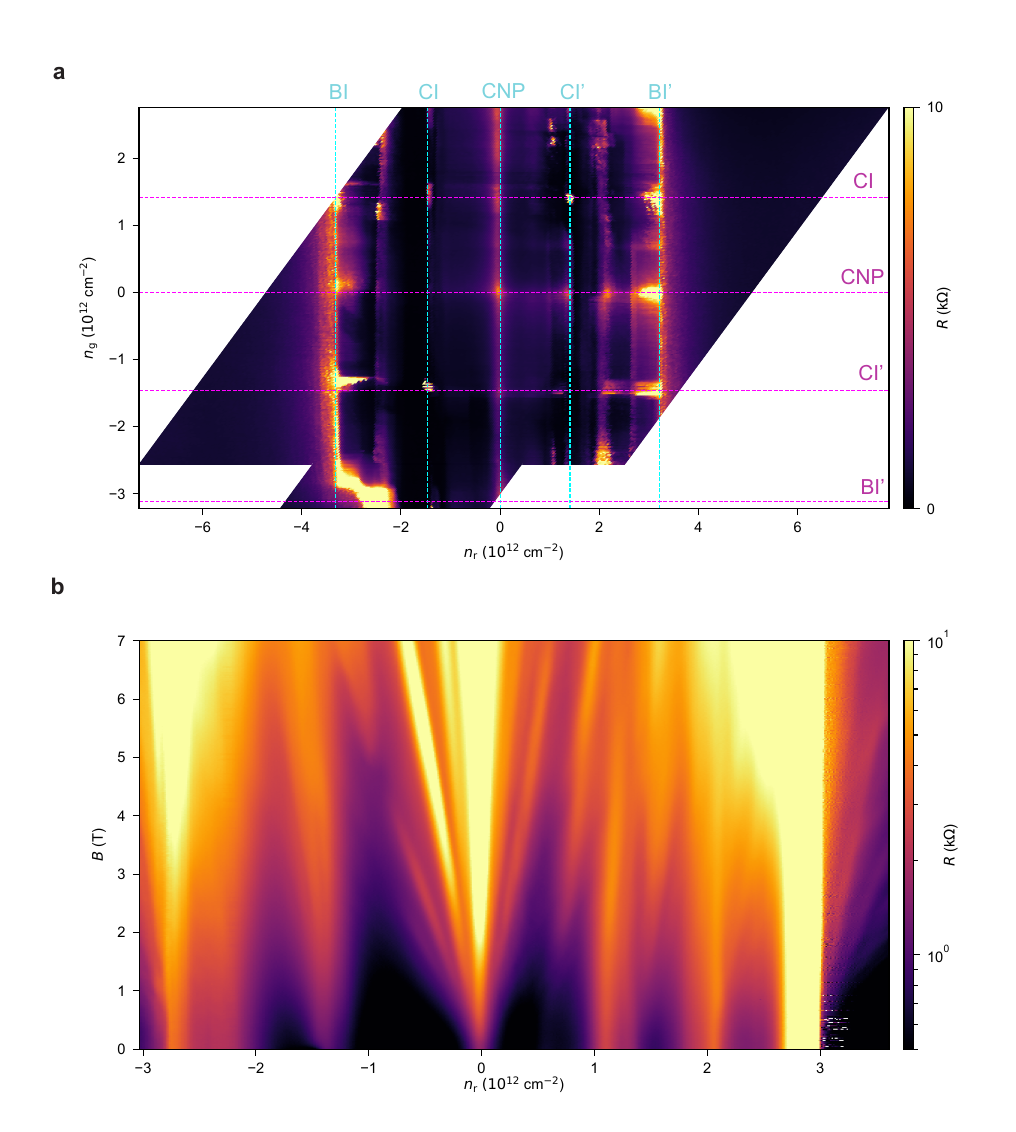}
\caption{\textbf{a} Resistance ($R$) across the device as a function of the carrier density controlled by the back gate ($n_\textrm{g}$) and the top gate ($n_\textrm{r}$). Densities are calculated from the capacitance model described in the Methods section. The dashed lines indicate the density values at which the charge neutrality point (CNP), CI and BI states arise in the region solely controlled by the back gate (pink) and in the top gated region (blue), both in the electron (with apostrophe) and hole (without) side. We use the BI densities to extract the twist angle of the device. \textbf{b} Magnetic field and $n_{\textrm{r}}$ dependence of the resistance across the device (Landau fan) at $V_\textrm{rg} = \SI{0}{V}$. The measurement is taken at 2 K. Both measurements are taken using an AC current of 1 nA.}
\label{fig:S_D}
\end{figure}

\clearpage
\newpage


\subsection{Critical density oscillations at different DC-bias currents and temperatures}
\begin{figure}[h!]
\centering
\includegraphics{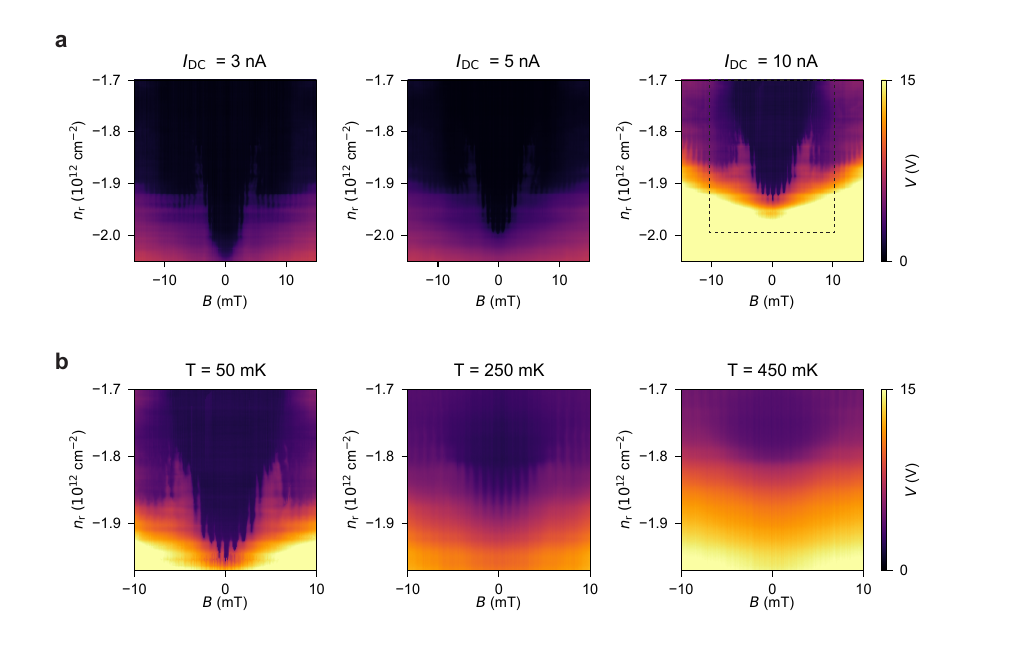}
\caption{\textbf{a} Critical density oscillations for $I_\textrm{DC} = 3, 5, 10$ nA at 50 mK. For low $I_\textrm{DC}$ (3 nA), the oscillations are not visible at around zero magnetic field. Increasing $I_\textrm{DC}$, oscillations appear starting at zero magnetic field (5 and 10 nA), and the onset density $n_\textrm{r}$ of the low resistance region moves towards more negative values. The dashed square in the top right colormap ($I_\textrm{DC} = 10$ nA) indicates the magnetic field and density range of the maps in \textbf{b}. \textbf{b} Temperature dependence of the critical density oscillations with $I_\textrm{DC} = 10$ nA. Increasing temperature, the extent in $n_\textrm{r}$ of the low-resistance region shrinks, and the amplitude of the oscillations increases. The oscillations vanish out at 450 mK.} 
\label{fig:S}
\end{figure}

\clearpage
\newpage


\subsection{Critical current oscillations in the CI regime}
\begin{figure}[h!]
\centering
\includegraphics{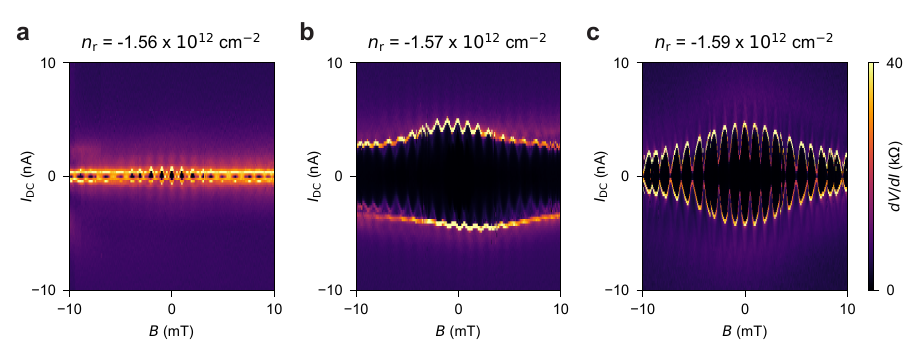}
\caption{\textbf{a-c} Critical current oscillations in the (SC, CI) regime, for the ring densities $n_\textrm{r} = -1.56 \times 10^{12} \SI{}{cm}^{-2}, -1.57 \times 10^{12} \SI{}{cm}^{-2}, -1.59 \times 10^{12} \SI{}{cm}^{-2}$. The same colorbar applies to the three maps. The oscillation period (peak-to-peak distance of critical current) is 1.13 mT and independent of $n_\textrm{r}$. At a low density ($n_\textrm{r} = -1.56 \times 10^{12} \SI{}{cm}^{-2}$), a chain of low resistance (superconducting) region appears in between the high resistance (normal conducting) background. The appearance of the oscillations drastically changes as the density is slightly modulated to -1.57 and -1.59 $\times 10^{12} \SI{}{cm}^{-2}$. For those densities, oscillations in resistance are no longer observed for $I_\textrm{DC}=\SI{0}{nA}$, and only the critical current oscillations appear. We also note that the pattern at $n_\textrm{r} = -1.57 \times 10^{12} \SI{}{cm}^{-2}$ is largely tilted.}
\label{fig:S}
\end{figure}

\clearpage
\newpage


\subsection{Critical current, critical field and critical temperature}
\begin{figure}[h]
\centering
\includegraphics{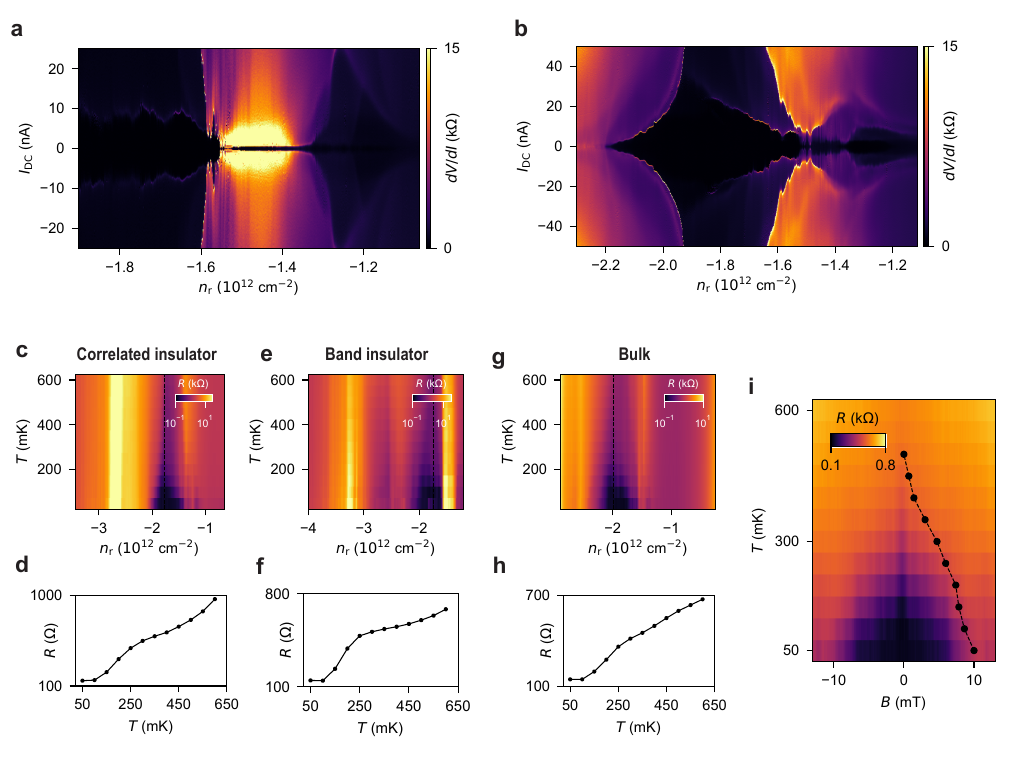}
\caption{\textbf{a}, \textbf{b}: Differential resistance $dV/dI$ as a function of $n_\textrm{r}$ when $n_\textrm{g} = -1.49 \times 10^{12}\SI{}{\cm }^{-2}$ (\textbf{a}, in the CI state) and $n_\textrm{g} = -3.13 \times 10^{12}\SI{}{\cm }^{-2}$ (\textbf{b}, in the BI state). \textbf{c}, \textbf{e}, \textbf{g}: Phase diagrams of the sample as a function of temperature and carrier density. Three configurations are investigated: \textbf{c} the ring surrounded by CI ($n_\textrm{g}=-1.49 \times 10^{12}\SI{}{\cm }^{-2}$), \textbf{e} the ring surrounded by BI ($n_\textrm{g}=-3.13 \times 10^{12}\SI{}{\cm }^{-2}$) and \textbf{g} bulk ($V_\textrm{rg}=0$). (\textbf{d}, \textbf{f}, \textbf{h}): Temperature dependence of the resistance at a fixed $n_\textrm{g}$, in the superconducting dome. The dashed lines in \textbf{c}, \textbf{e}, \textbf{g} show the point at which the data is taken. \textbf{i} Temperature and magnetic field dependence of the resistance when $n_\textrm{g}=-2.02 \times 10^{12}\SI{}{\cm }^{-2}$ ($V_\textrm{rg} = 0$ V). The markers show the critical magnetic field $B_\textrm{c}$ at which the resistance increases. By fitting the temperature $T$ dependence of $B_\textrm{c}$ with $B_\textrm{c}(T)=\frac{\phi_0}{\xi^{2}_\textrm{BCS}}(1-\frac{T}{T_\textrm{c}})$, we estimate the BCS coherence length $\xi_\textrm{BCS}$ to be around 160 nm.
All the measurements are taken using an AC current of 1 nA. }
\label{TcHc}
\end{figure}


\clearpage
\newpage


\subsection{Spectrogram of quantum oscillations in the BI regime}
\begin{figure}[h]
\centering
\includegraphics{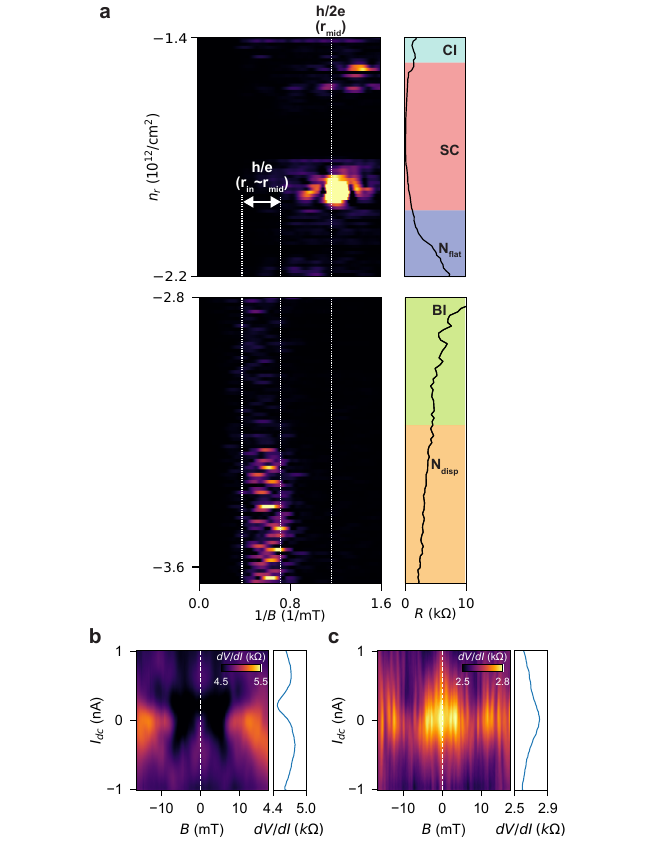}
\caption{\textbf{a} Spectrum of the magneto-resistance oscillation with the ring surrounded by BI state. The white-dashed line labelled $h/2e$ ($r_\textrm{mid}$) indicates the frequency corresponding to an $h/2e$-periodicity with $r_\textrm{eff}=r_\textrm{mid}$. The frequency range labelled $h/e$ ($r_\textrm{in} \sim r_\textrm{mid}$), indicates the region with $h/e$-periodicity with $r_\textrm{eff}$ between $r_\textrm{in}$ and $r_\textrm{mid}$. \textbf{b} Differential resistance as a function of $I_\textrm{DC}$ and $B$ for the ($N_\textrm{flat}$, BI) regime (left) and the ($N_\textrm{disp}$, BI) regime (right). The white dashed line indicates $B=0$. In the side panel of each map, $dV/dI$ is plotted as a function of $I_\textrm{DC}$ at $B=0$.}
\label{Supp_ABinBI}
\end{figure}

\clearpage
\newpage

\subsection{Electrostatic simulation}

\begin{figure}[h]
\centering
\includegraphics{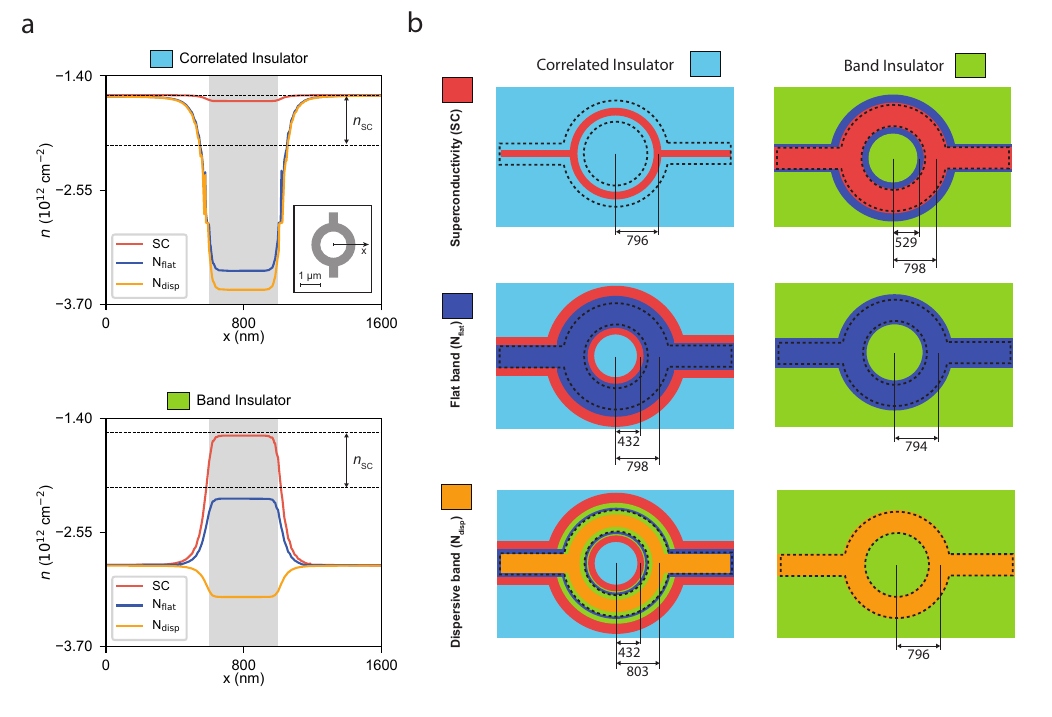}
\caption{Electrostatic simulations of the spatial distribution of the carrier density in the ring. \textbf{a} Carrier density distribution along the radial ($x$) axis. The inset in the top panel shows schematics of the structure and indicates the direction of the $x$ axis. In the top panel, the bulk is set in the correlated insulator (CI) state and in the bottom panel, in the band insulator (BI) state. For CI and BI cases, we simulate the density distribution with the ring in the superconducting (SC), normal conducting flat band ($N_\textrm{flat}$) or dispersive band ($N_\textrm{disp}$). The gray shaded area indicates the lithographic width of the ring (400 nm), and the horizontal dotted lines are the densities at which the ring is superconducting ($n_\textrm{sc}$). \textbf{b} Spatial density distribution and stripe structure of the ring for the six regimes. Numbers with arrows indicate the dimension of each part, in the units of nm.}
\label{fig:simulation}
\end{figure}

We perform electrostatic simulations of the carrier density distribution in the MATBG ring presented in this work. The inset in the top panel of Fig. \ref{fig:simulation}a shows a top view of the geometry of the structure. 
In the simulations, we discretize the system with a finite element model and self-consistently solve the Poisson equation ($\varepsilon_0 \varepsilon_r (\textbf{x}) \nabla V(\textbf{x}) = \rho(\textbf{x})$) using the software COMSOL Multiphysics. In the equation, $\varepsilon_0$ is vacuum permittivity, $\varepsilon_\textrm{r}$ is relative permittivity of the dielectric layer ($\varepsilon_\textrm{r}$(hBN) = 3.3 and $\varepsilon_\textrm{r}$(AlOx) = 9.5), $V(\textbf{x})$ is electric potential at position $\textbf{x}$, and $\rho(\textbf{x})$ is the charge density at position $\textbf{x}$. 
We impose $\rho(\textbf{x}) = 0$ in the dielectric layers and implement $\rho(\textbf{x})$ in the MATBG plane using the Thomas-Fermi approximation \cite{Luscombe_1992}. We simplify the band structure of MATBG assuming parabolic bands for both the flat and the dispersive bands, with respective effective masses $m_\textrm{flat} = 0.05m_\textrm{e}$ and $m_\textrm{disp} = 0.4m_\textrm{e}$ \cite{cao_unconventional_2018,rodan-legrain_highly_2021}. We assume a band gap of 30 meV for the BI state, separating the flat and dispersive bands \cite{rodan-legrain_highly_2021}. We impose $V = V_\textrm{rg}$ on the top side of the top hBN layer and $V = V_\textrm{bg}$ at the bottom of the bottom hBN layer. We approximate the CI state with a smaller energy gap of 0.3 meV. After solving Poisson's equation, the carrier density $n(\textbf{x})$ is directly extracted from the calculated surface charge density $\rho_\textrm{S} (\textbf{x})$ using $\rho_\textrm{S} (\textbf{x})= -e n (\textbf{x})$.

The simulated carrier density distributions along the horizontal cut across the arm of the ring are shown in Fig. \ref{fig:simulation}a. The left and right panels correspond, respectively, to the regimes in which the outside of the ring is in the CI ($V_\textrm{bg} = -4.5$ V) and BI regime ($V_\textrm{bg} = -8.2$ V). In both panels, we show the charge distribution when the ring is set to the superconducting state ($V_\textrm{rg} = \SI{-0.065}{V}$ when CI outside and $V_\textrm{rg} = \SI{1.69}{V}$ when BI outside), normal conducting in the flat band ($V_\textrm{rg} = \SI{-2.31}{V}$ when CI outside and $V_\textrm{rg} = \SI{0.88}{V}$ when BI outside) and in the dispersive band ($V_\textrm{rg} = \SI{-2.55}{V}$ when CI outside and $V_\textrm{rg} = \SI{-0.44}{V}$ when BI outside). 
The vertical dashed lines indicate the lithographic width of the ring (400 nm), while the horizontal lines show the density range of the superconducting state. For all six conditions, the fringing field results in a broadening of the carrier density distribution, making the actual width of the targeted quantum state wider/smaller than the lithographic width of the ring gate.
Fig. \ref{fig:simulation}b shows the carrier density distribution in the MATBG layer for the six configurations. Because the quantum states in MATBG are density-dependent, the broadening of the carrier density distribution leads to a stripe structure of phases across the sample. The insets in Fig. \ref{fig:simulation}b. show the quantum state distribution across the arm of the ring. The charge densities at which each quantum state (CI, SC, N$\textrm{flat}$) arises are extracted from the map in Fig. \ref{fig:S_D}a. Owing to the circular symmetry of the system, multiple quantum state stripes form a ring with different radii and could contribute to oscillations.

Even in (N$_\textrm{flat}$, CI) regime, a superconducting path can be formed. In particular, there is a small superconducting ring with a radius of 432 nm and an open path surrounding the outer edge of the N$_\textrm{flat}$ region. Such a spurious superconducting path could contribute to the transport characteristics such as the one observed in Fig. \ref{Nflat}. Moreover, the oscillations in resistance and critical current observed in the (N$_\textrm{flat}$, CI) regime have a period of 3.19 mT, corresponding to an effective radius of 446 nm in $h/2e$.
This number coincides with the small superconducting ring predicted by the simulation, though more work is necessary to confirm the existence of such a ring by taking the superconducting proximity effect into account.

Such a satellite superconducting path is also predicted in (N$_\textrm{disp}$, CI) regime, where only the conventional Aharonov--Bohm effect ($r_\textrm{eff}$ in between $r_\textrm{in}$ and $r_\textrm{mid}$) and no superconducting-like nonlinearity are experimentally observed. In this regime, there is a BI state between the N$_\textrm{disp}$ state and the satellite superconducting path, which might make the majority of the current flow away from the superconducting path.

\newpage
\subsection{Tunable phase shift of the Little--Parks effect}
\begin{figure}[h]
\centering
\includegraphics{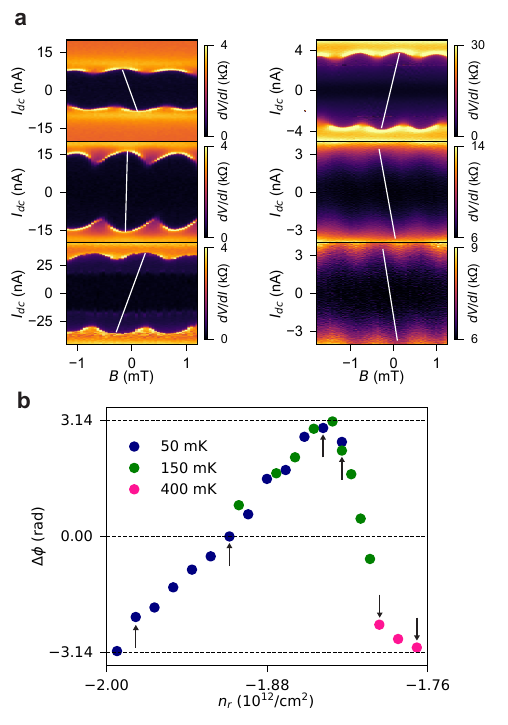}
\caption{ \textbf{a} Detailed evolution of the asymmetry with $n_\textrm{r}$. The white lines connect the maximum critical current at positive and negative $I_\textrm{DC}$. \textbf{b} Phase-shift $\Delta \Phi$ as a function of $n_\textrm{r}$ at 50, 150 and 400 mK.
}
\label{PhaseShift}
\end{figure}

The magnetic field at which the critical current reaches its maximum ($B_\textrm{max}$) depends on the sign of the current, as can be seen in Fig.~\ref{PhaseShift}a. This means that there is a phase shift in the magneto-oscillation of the critical current depending on the polarity of the bias current. Here, we show that the ring gate tunes this phase shift.
To characterize this phase shift, we extract $\Delta B=B_\textrm{max}(I>0)-B_\textrm{max}(I<0)$ and convert it to a phase shift in flux $\Delta\phi=\frac{\Delta B\times\pi (R_\textrm{eff})^{2}}{h/2e}$.
As shown in Fig. \ref{PhaseShift}.b, $\Delta\phi$ monotonically decreases from $\pi$ to $-\pi$ and vanishes when $n_\textrm{r} = -1.90 \times 10^{12} \SI{}{cm}^{-2}$.

Such an asymmetry can be attributed to the slight shape-asymmetry of the superconducting path inside the ring, which is presumably gate-dependent. It could result in a difference in the inductance of each arm.

Our observations imply that special care has to be taken when analyzing the phase of the Little--Parks oscillations in critical current in a thin superconductors, as the large inductance of the material can result in an inevitable phase shift. 

\newpage

\subsection{Interpretation of the temperature dependence of the Little--Parks effect}
\begin{figure}[h]
\centering
\includegraphics{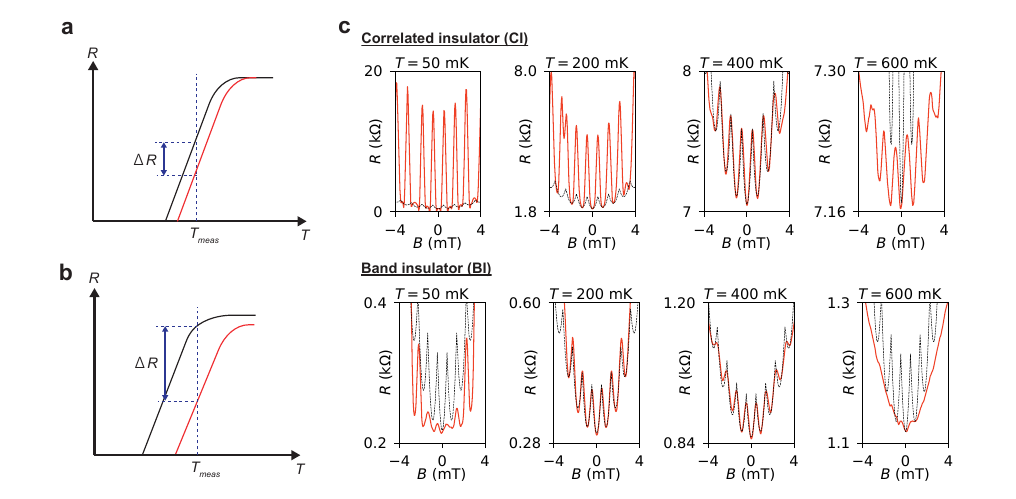}
\caption{Illustration of the temperature dependence of resistance of a ring with (black) and without (orange) magnetic field. \textbf{a} The ideal case where the shape of the temperature dependence does not depend on the magnetic field, namely R(T). \textbf{b} The case with magnetic field dependence, namely R(T,B). \textbf{c}: Results of the fitting of the data to Tinkhams's formula for the (SC, CI) regime (top panel) and (SC, BI) regime (bottom panel).}
\label{fig:LPTdep}
\end{figure}

The Little--Parks effect is usually measured as a magneto-resistance oscillation and then converted to an oscillation of the critical temperature. There, a translation of $\Delta R=\frac{dR}{dT}\times\Delta T_\textrm{c}$ is used (see Fig. \ref{fig:LPTdep}(a). Here, $\Delta R$ and $\Delta T_\textrm{c}$ are the amplitude of the magneto oscillation of the resistance and the critical temperature, respectively.
Tinkham derived the theoretical formula for $\frac{\Delta T_\textrm{c}}{T_\textrm{c}}=\frac{0.55(\xi_0)^2}{R^2}(N-\frac{\phi}{\phi_0})^2$, where $T_\textrm{c}$ is the critical temperature at zero magnetic field, $\xi_0$ is the BCS coherence length, and $N$ an integer number. Here, it is assumed that the shape of the temperature dependence of the resistance $R(T)$ does not change with the magnetic field, making $\Delta R$ constant in a wide range of temperature as shown in \ref{fig:LPTdep}(a).

However, as shown in Fig. \ref{fig:LPTdep}, the Little--Parks effect observed here strongly depends on the temperature. In addition, as we show here, fitting the magneto-resistance data to Tinkham's formula is valid in a limited range of temperatures even though the observed amplitude of $\Delta R$ is larger than the expected value by a factor of ten.
Such discrepancies could arise from the breakdown of the assumption that the shape of the temperature dependence of the resistance does not change with the magnetic field. In 2D superconductors such as MATBG, the thickness is always shorter than the penetration length of the magnetic field (Pearl length), letting magnetic flux always penetrate as soon as the magnetic field is applied. Therefore, the critical temperature is modulated not only by the Little--Parks effect but with the magnetic field itself, making Tinkham's assumption break. 

Also, we discuss that the observation of the Little--Parks effect prefers a temperature that is close enough to the critical temperature.
The condensation energy $U$ of a superconductor with a width $w$, thickness $t$ and length $L$ is $U=\frac{1}{2\mu_0}(\frac{\phi_0}{2\pi\xi^2})^{2}Lwt$. Here, $\mu_0$ is vacuum permeability, and $\xi$ is the Ginzburg-Landau coherence length $\xi=\frac{\xi_0}{(1-\frac{T}{T_\textrm{c}})^{1/2}}$ with $\xi_0$ being the zero-temperature coherence length. Then, $U<k_BT$ sets an approximate condition that such a piece of superconductor becomes unstable. At a fixed temperature $T$, the maximum length of such a broken area is
\begin{equation}
L=\frac{2\mu_0k_BT}{wt}\left(\frac{2\pi\xi^{2}}{\phi_0}\right)^{2}.
\label{length}
\end{equation}

Then, if $L<2\pi R$, the superconductor hosts a piece of normal conducting part, which does not extend across the entire ring. Such a part would make a Josephson junction in the system. Equation \ref{length} leads to the condition that
\begin{equation}
    k_BT\leq \frac{Rwt\phi_0^2}{4\pi\mu_0\xi^{4}}=\frac{Rwt\phi_0^2}{4\pi\mu_0\xi_{0}^{4}}\left(1-\frac{T}{T_\textrm{c}}\right)^{2}.
\label{condition}
\end{equation}


Equation \ref{condition} predicts a characteristic temperature $T^{*}$ at which the non-superconducting regime extends the entire ring. $T^{*}$ is analytically obtained
\begin{equation}
T^{*}=\frac{T_\textrm{c}(-\sqrt{T_\textrm{c}(T_\textrm{c}+4\alpha)}+2\alpha+T_\textrm{c})}{2\alpha}.
\label{Tstar}
\end{equation}
Here, $\alpha=\frac{Rwt\phi_0^2}{4\pi\mu_0\xi_{0}^{4}}$.
It can be seen that $T^{*}$ is very close to $T_\textrm{c}$. For example, if $R=850$ nm, $\xi_0=160$ nm, $w=400$ nm, $t=0.34$ nm, and $T_\textrm{c}=600$ mK, $T^{*}=594$ mK. Though the above calculation is a crude estimation, it is suggested that the ring becomes more homogeneous as the temperature increases close to the critical temperature. This supports that the observed asymmetry in the Little--Parks oscillations becomes less prominent at higher temperatures and with a density $n_\textrm{r}$ where the superconducting state has a small critical current.

\subsection{Interpretation of the beating pattern}
As shown in Fig \ref{TunableLP}, the critical current reaches zero at $\Delta B\sim\SI{5}{mT}$ in the (SC, BI) regime. One explanation of such an effect is the formation of a spurious Josephson junction in the system, that would produce a Fraunhofer-like pattern. However, this hypothesis fails to explain the observed data quantitatively. Namely, for a 2D planar Josephson junction \cite{clem_josephson_2010,fermin_beyond_2023}, the magnetic field at which the critical current vanishes is $\sim 1.8(h/2e)/W^{2}$, where $W$ is the lateral width of the junction. 
Using this formula, the magnetic field of $\Delta B\sim\SI{5}{mT}$ is converted to an effective junction width of $W\simeq\SI{860}{nm}$. This dimension significantly exceeds the width of the arm of the ring (400 nm), inferring that the superconductivity inside the ring is homogeneous and does not host any junction.

Another possibility that could explain the vanishing critical current is the existence and interference of multiple superconducting paths inside the ring. In fact, the central frequency of the Little--Parks oscillation ($\sim$1.1/mT) coincides with the center radius of the ring ($\sim$855 nm). If superconducting paths with effective radii of $\sim$87 nm (frequency $\sim$0.8/mT) or $\sim$650 nm (frequency $\sim$1.2/mT) are also present, their interference would produce a beating frequency of 0.2/mT that equals to the beating periodicity of 5 mT.

\subsection{Distinguishing Little--Parks effect from SQUID}
\begin{figure*}[!h]
\centering
\includegraphics[width=1\textwidth]{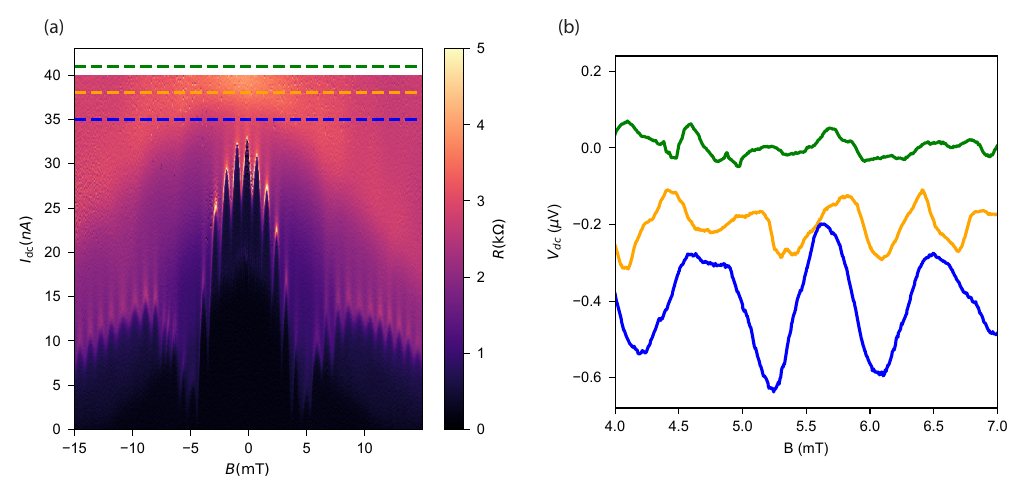}
\caption{\textbf{a} Resistance across the device as a function of magnetic field and current bias. The colored dashed lines correspond to the traces shown in Figure (b). \textbf{b} Line cuts of Figure a where the voltage drop across the device is shown as a function of magnetic field at different current biases. The value of the top gate voltage at which this data was taken is $V_\textrm{rg} = \SI{1.30}{V}$}
\label{fig:S_A}
\end{figure*}

\begin{figure*}[!h]
\centering
\includegraphics[width=1\textwidth]{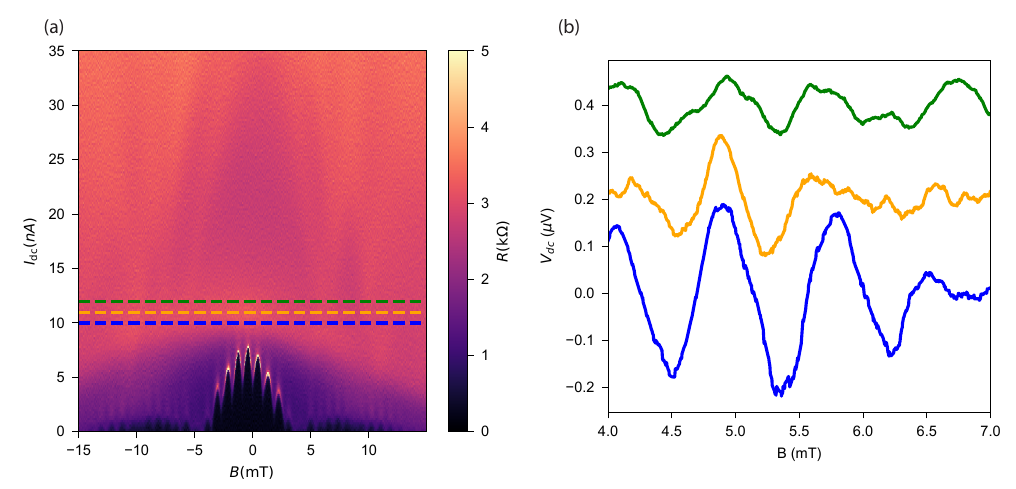}
\caption{\textbf{a} Resistance across the device as a function of magnetic field and current bias. The colored dashed lines correspond to the traces shown in Figure (b). \textbf{b} Line cuts of Figure (a) where the voltage drop across the device is shown as a function of magnetic field at different current biases. The value of the top gate voltage at which this data was taken is $V_\textrm{rg} = \SI{1.14}{V}$}
\label{fig:S_B}
\end{figure*}

\begin{figure*}[!h]
\centering
\includegraphics[width=1\textwidth]{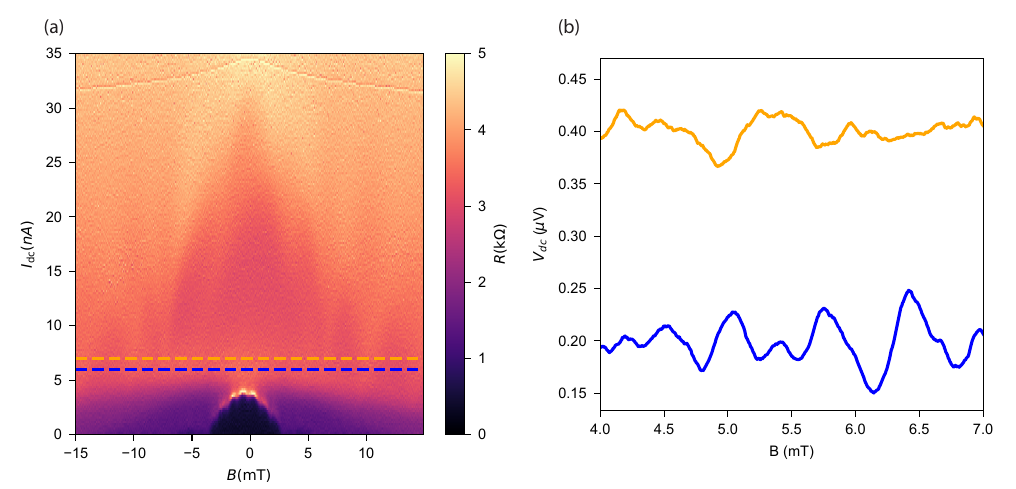}
\caption{\textbf{a} Resistance across the device as a function of magnetic field and current bias. The colored dashed lines correspond to the traces shown in Figure (b). \textbf{b} Line cuts of Figure (a) where the voltage drop across the device is shown as a function of magnetic field at different current biases. The value of the top gate voltage at which this data was taken is $V_\textrm{rg} = \SI{1.05}{V}$}
\label{fig:S_C}
\end{figure*}

Due to the fluxoid quantization in superconductors \cite{tinkham2004introduction}, when a magnetic field is applied perpendicular to the device plane, there will be a circulating current around any hole threaded by the field. If that current flows in a region of the superconductor that has a nonzero resistance this will lead to the so-called Little--Parks effect \cite{little_observation_1962,groff_fluxoid_1968}. In the case of 3D superconductors, current flows through the bulk, without encountering any resistance. However, for two-dimensional materials, where the penetration depth of the magnetic field is bigger than the thickness of the material, the whole superconducting surface has nonzero resistance. The Little--Parks effect is thus expected in any two-dimensional superconducting device. 
\\
\\
The question that remains in the case of a two-dimensional ring is whether there can be spontaneous formation of Josephson junctions leading to a SQUID-like behaviour. To discern between the two behaviours, we focus on the oscillations in voltage drop across the device when biased \textit{above} the critical current. There, oscillations in the voltage drop are expected both in the Little--Parks and SQUID cases. In the former, they are expected due to the finite breadth in current of the superconducting transition. In the case of the latter, the oscillations are expected due to the fact that even if the junctions present in the ring are biased above their critical currents, the superconducting condensate in the rest of the ring is still present and thus superconducting coherence and subsequent interference phenomena still take place \cite{clarke2006squid}.
\\
\\
The difference between the two types of oscillations is that for the SQUID ones a particular shape is expected while the ones stemming from Little--Parks effect are not expected to take any particular shape. For the ones originating from SQUID behaviour, the following voltage dependence is expected in the case of two overdamped junctions:

\begin{equation}
    V = (R/2) \sqrt{I^2 - (2I_\textrm{c} \cos{(\pi \Phi / \Phi_0 ))^2}}
\end{equation}

where $R/2$ is the resistance of the two Josephson junctions in parallel, $\Phi$ the flux threading the loop, and $\Phi_0$ the superconducting flux quantum. Assuming that the eventual junctions that form in the material are overdamped is reasonable taking into account the low critical currents and quasiparticle resistances expected in junctions in this material, from measurements performed in devices with electrostatically engineered junctions \cite{portoles_tunable_2022}. Figure \ref{fig:S_D} shows the expected voltage drop for such a case. Figures \ref{fig:S_A}-\ref{fig:S_C} show the corresponding line plots taken at different current biases when sweeping the magnetic field. We observe no particular resemblance between the data and the expected trend. As a comparison, we performed the same analysis for a SQUID device in this material \cite{portoles_tunable_2022}, where the resemblance was clear. Thus, we conclude that it is safe to consider the oscillations as stemming mostly from the Little--Parks effect.


\begin{figure*}[!h]
\centering
\includegraphics[width=0.8\textwidth]{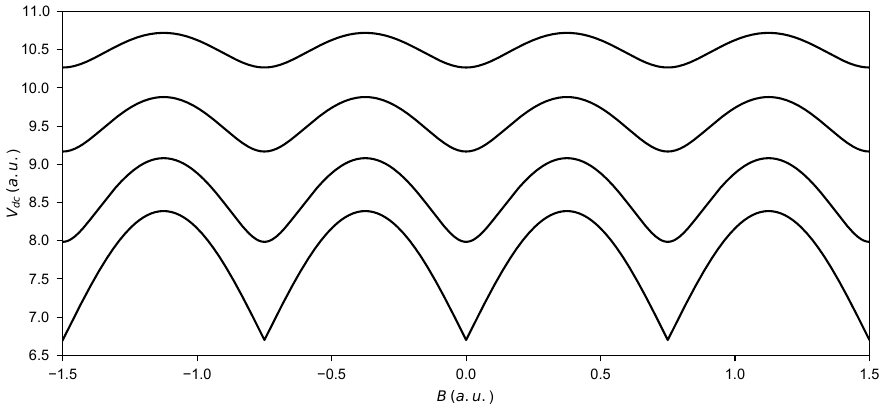}
\caption{Expected voltage drop across a SQUID device with overdamped Josephson junctions as a function of magnetic field. Each solid line corresponds to a different bias current. There is an arbitrary offset for visibility purposes. A higher offset corresponds to a higher current bias.}
\label{fig:S_D}
\end{figure*}

\clearpage
\newpage

\bibliographystyle{apsrev4-1}
\bibliography{Bibliography.bib}

\end{document}